# Band structure of hydrogenated silicene on Ag(111): Evidence for silicane

W. Wang*, W. Olovsson and R. I. G. Uhrberg

*Department of Physics, Chemistry, and Biology, Linköping University, S-581 83 Linköping, Sweden*

In the case of graphene, hydrogenation removes the conductivity due to the bands forming the Dirac cone by opening up a band gap. This type of chemical functionalization is of utmost importance for electronic applications. As predicted by theoretical studies, a similar change of the band structure is expected for silicene, the closest analogue to graphene. We here report a study of the atomic and electronic structures of hydrogenated silicene, so called silicane. The ($2\sqrt{3}\times2\sqrt{3}$) phase of silicene on Ag(111) was used in this study since it can be formed homogeneously across the entire surface of the Ag substrate. Low energy electron diffraction and scanning tunneling microscopy data clearly show that hydrogenation changes the structure of silicene on Ag(111) resulting in a (1×1) periodicity with respect to the silicene lattice. The hydrogenated silicene also exhibits a quasi-regular ($2\sqrt{3}\times2\sqrt{3}$)-like arrangement of vacancies. Angle resolved photoelectron spectroscopy revealed two dispersive bands which can be unambiguously assigned to silicane. The common top of these bands is located at ~0.9 eV below the Fermi level. We find that the experimental bands are closely reproduced by the theoretical band structure of free standing silicane with H adsorbed on the upper hexagonal sub-lattice of silicene.

Silicene is, in analogy with graphene, predicted to have a band structure forming a Dirac cone (occupied by massless fermions) with the apex at the Fermi level. Two dimensional (2D) materials with this kind of exceptional band structure are of great interest for various electronic applications [1-3]. This requires, however, different kinds of functionalization. Adsorption of hydrogen atoms is one method which is of great interest since theoretical studies show that it leads to an opening of a band gap which is crucial for applications in future nano-electronic devices [4]. An experimental determination of the band structure of hydrogenated silicene, named silicane, is therefore of utmost importance. Silicene is mostly synthesized on Ag(111) substrates because of a low Si-Ag intermixing and a "matching" 3 to 4 ratio between the lattice constants of silicene and the Ag(111) surface [5]. After many years of silicene studies, experimental evidence of a Dirac cone is still missing. As suggested by theoretical calculations, the absence of Dirac fermions is due to strong hybridization between Si and Ag atoms [6]. Even though, Dirac cones have not been experimentally observed, a silicene field-effect transistor was recently reported using (4×4) and ($2\sqrt{3}\times2\sqrt{3}$) silicene grown on a Ag(111) thin film on mica [7]. This publication has shown the great potential of silicene in nano-electronic devices. This emphasizes the importance of functionalization of graphene-like 2D materials, a topic which we address here by reporting the atomic and electronic structure of silicane.

In this paper, silicane formed by hydrogenation of ($2\sqrt{3}\times2\sqrt{3}$) silicene on Ag(111) has been studied by low energy electron diffraction (LEED), scanning tunneling microscopy (STM), and angle-resolved photoelectron spectroscopy (ARPES). LEED patterns show that the hydrogenation changes the structure of the silicene sheet to essentially a (1×1) structure with some higher order periodicity indicated by weak additional diffraction spots. STM images of the silicane reveal a hexagonal structure with the periodicity of the silicene (1×1) lattice and vacancies with a quasi ($2\sqrt{3}\times2\sqrt{3}$) periodicity. The ARPES data show two dispersive bands around normal emission which are in good agreement with the theoretical band structure, calculated for a silicane model with hydrogen adsorbed on one of the two hexagonal sub-lattices of free-standing silicene.

Samples were prepared in ultrahigh vacuum (UHV) according to the following procedure. A clean Ag(111) surface was prepared by repeated cycles of sputtering by Ar$^+$ ions (1 keV) and annealing at approximately 400 °C until a sharp (1×1) LEED pattern was obtained and a sharp Shockley surface state was observed in ARPES. About 1 monolayer (ML) of Si was deposited at a rate of ~0.03 ML/min while the Ag(111) substrate was kept between 280 and 300 °C. This type of preparation repeatedly results in a full coverage of ("$2\sqrt{3}\times2\sqrt{3}$") silicene on Ag(111), as reported in Ref. [8]. We use the notation ("$2\sqrt{3}\times2\sqrt{3}$") to indicate that this silicene phase has an atomic structure that deviates from a simple ($2\sqrt{3}\times2\sqrt{3}$) periodicity, as discussed in Refs. [8-10]. After a quality check by LEED and STM, the ("$2\sqrt{3}\times2\sqrt{3}$") silicene was exposed to atomic hydrogen generated by dissociating H$_2$ molecules at a hot W filament (1800 °C) placed ~ 6 cm in front of the sample. The exposures were quantified during the experiments in terms of the number of Langmuirs (L) of H$_2$ molecules (1 L = 1×10$^{-6}$ Torr s). STM images were recorded at room temperature using an Omicron variable temperature STM in a UHV system at Linköping University which was also equipped with a LEED optics. ARPES data and LEED patterns were obtained at the MAX-lab synchrotron radiation facility using the end station at beam line I4. ARPES data were acquired at room temperature by a Phoibos 100 analyzer from Specs with a two-dimensional detector. The energy and angular resolutions were 50 meV and 0.3°, respectively. First-principles density functional theory (DFT) calculations were performed using the full potential linearized augmented plane wave method as implemented in the WIEN2k package [11]. The positions of the atoms were fully relaxed using the projector augmented wave method (PAW) Vienna ab initio simulation package (VASP) code [12]. For free standing silicane, the energy cutoff of the plane-wave basis set was 375 eV, and the k-point mesh was 9×9×1. All calculations employed the generalized gradient approximation (GGA) for exchange and correlation, according to Perdew *et al.* [13].

*Corresponding author: weiwa49@ifm.liu.se

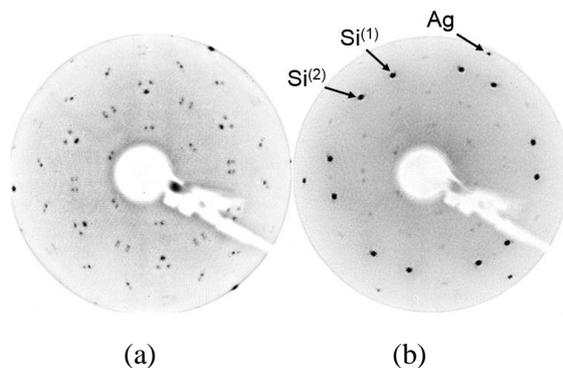

Fig.1 (a) LEED pattern (77 eV) of ("2√3×2√3") silicene on Ag(111) showing diffraction from the two possible silicene orientations rotated by +10.9 and -10.9° compared to Ag(111), respectively. Some of the diffraction spots originate from the moiré-like long range order observed by STM, see Fig. 2(a). For a detailed assignment of the diffraction spots see Ref. [8]. (b) LEED pattern (82 eV) of the ("2√3×2√3") silicene surface exposed to atomic hydrogen. The arrows indicate an Ag 1×1 diffraction spot and two diffraction spots, $Si^{(1)}$ and $Si^{(2)}$, from the two rotated silicene sheets, respectively.

Silicene synthesized on Ag(111) can form various phases, such as (4 × 4), (√13×√13) and (2√3×2√3). So far, it has only been experimentally verified that one of them, the (2√3×2√3) phase, here denoted as ("2√3×2√3"), can cover the whole surface without the coexistence of any other phases [8,14]. Hence, the ("2√3×2√3") phase is the best choice for a study of the properties of hydrogenated silicene. Before hydrogenation, the quality of the as prepared ("2√3×2√3") silicene was confirmed by LEED and STM to match the silicene in our previous publication [8]. Fig. 1(a) shows a typical LEED pattern of ("2√3×2√3") silicene on Ag(111). During hydrogenation, there is a charge transfer occurring from Si atoms to H atoms. The hybridization changes from mixed $sp^2/sp^3$ to $sp^3$ hybridization [16]. After exposure to atomic hydrogen, the LEED pattern in Fig. 1(b) shows clear changes. Some of the higher order diffraction spots are absent while the other are significantly weaker. In contrast, the silicene (1×1) spots are brighter. The LEED pattern in Fig. 1(b) is dominated by diffraction from the two differently oriented silicene sheets, indicated by $Si^{(1)}$ and $Si^{(2)}$. In a subsequent investigation, we found that the quenched and the diminished spots can be brought back by annealing at 270 °C for a short time (~5 mins), which shows that the process of hydrogenation on silicene is more easily reversible than on graphene [17]. After an extended exposure (~6 times saturation exposure) the LEED pattern shows only (1×1) Ag spots (not shown in the paper) and annealing cannot bring back diffraction spots from silicene anymore. Based on a theoretical study, hydrogenation of the silicene layer on Ag(111) is a self-limited process [18]. A H-coverage of 50 % was reported to be energetically favorable. For continued hydrogenation, the hydrogen atoms were predicted to adsorb on both sides of the silicene layer, leading to a detachment of the silicene layer from the Ag(111) surface. This could explain why there was no sign of any Si related diffraction after the extended hydrogen exposure.

Information about the morphology of the initial silicene and the silicane formed by hydrogenation was obtained by STM. The atomically resolved filled state STM image in Fig. 2(a) shows the structure of ("2√3×2√3") silicene on Ag(111) with a typical moiré-like long range pattern of bright honeycombs. Between the bright regular honeycombs there are distorted honeycombs. Both the regular and distorted honeycombs have very consistently been reported in the literature for ("2√3×2√3") silicene. Our previous work [8] confirms that this silicene phase covers the whole surface with a specific buckling for which only two out of the fourteen Si atoms per (2√3×2√3) unit cell are observed by STM. After hydrogenation, the structure is completely changed. The silicane shows a (1×1) periodicity with a quasi-periodic arrangement of defects as shown by Figs. 2(b) and 2(c). The green diamond corresponds to a silicene (1×1) periodicity and the red diamonds correspond to a (2√3×2√3) periodicity with respect to the Ag(111) lattice, i.e., (√7×√7)R19.1° with respect to silicene. The STM images indicate that every second Si atom of the silicene layer is bonding to a hydrogen atom resulting in the hexagonal structure. As a consequence of the hydrogenation, the complicated buckling of the initial ("2√3×2√3") silicene has been removed.

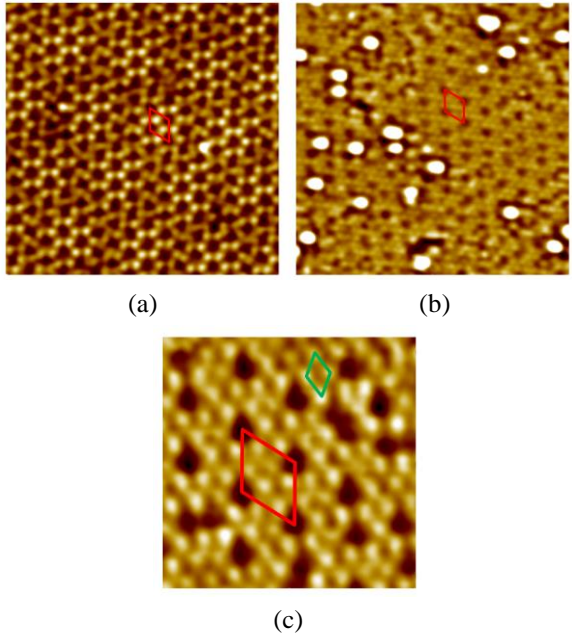

(a)  (b)

(c)

Fig. 2 (Color online) (a) Atomically resolved empty state STM images of a 14.2×14.2 nm² area of ("2√3×2√3") silicene on Ag(111). (b) A 14.2×14.2 nm² area of silicane formed by atomic hydrogen exposure (saturation exposure), showing a hexagonal pattern corresponding to a (1×1) periodicity of silicene and vacancies (dark) arranged in a quasi-periodic fashion. There are also larger bright protrusions in the STM image after the hydrogenation. (c) A 4.3×4.3 nm² area of hydrogenated silicene. The (1×1) periodicity is indicated by a green diamond and red diamonds illustrate the (2√3×2√3) cell with respect to Ag(111), i.e. (√7×√7)R19.1° with respect to silicene. The (2√3×2√3) cells in (b) and (c) describe the part where the vacancies are ordered. All STM images were recorded at room temperature in constant current mode with a tunneling current of 200 pA and a sample bias of +1.0 V for (a), and +50 mV for (b) and (c).

The defects, which appear darker than the hydrogen terminated Si atoms, are most naturally interpreted as vacancies. It is intriguing to find a quasi-periodic ordering of these vacancies. Locally, the defects are found to form a (2√3×2√3) structure as indicated by the red unit cell in Fig. 2(b). These vacancies are interpreted by Qiu *et al.* [14] as missing H atoms. In our study, we found that increasing hydrogen exposures do not decrease the density of the vacancies but start to remove Si atoms. After 1.33 times the saturation exposure, we found areas where Si had been removed showing bare Ag(111). An alternative explanation to the presence of vacancies could be that Si atoms are actually removed from the silicene sheet during hydrogenation. It is here interesting to note that our study and that of Qiu *et al.*, show bright protrusions of similar size and density (number/area). In Ref. [14] these protrusions were assigned to contamination during the H exposure. It is, however, not very likely that our and their experimental conditions would result in "identical" contamination effects. We therefore regard the bright protrusions as features that arise naturally from the hydrogenation process. A plausible explanation would be that the bright protrusions are small clusters formed by Si atoms that have been removed from the silicene layer, each leaving a vacancy behind (see the Supplemental Material [15]).

ARPES was performed along the $\bar{\Gamma} \to \bar{M}_{Ag}$ and $\bar{\Gamma} \to \bar{K}_{Ag}$ directions of the (1×1) SBZ of Ag(111), as illustrated by the black arrows in Fig. 3(a). Apart from dispersive features form the Ag(111) substrate (Ag-*sp* and U-*sp*), there are two clear bands, $S_1$ and $S_2$, which can be conclusively assigned silicane. They show good agreement with the calculated bands of free-standing silicane in Fig. 3(d), even though the band mapping directions are 10.9° off the high symmetry lines of the silicane SBZs. The directions used for the band mapping are equivalent for the two orientations of the silicane domains and they were chosen deliberately to avoid the complexity of two superimposed silicane band structures in the ARPES data. We also mapped the band structure along a direction corresponding to a symmetry line for one of the silicane orientations. Such a line corresponds to a path that is 21.8° off the symmetry direction for the other silicane orientation. The bands around normal emission did not show any noticeable difference but for large emission angles, the $S_1$ bands originating from each of the two silicane orientations were observed (see the Supplemental Material [15]). There are two differences between the ARPES data and the calculated band structure of free-standing silicane. One is that the calculation locates $\Sigma_1$ and $\Sigma_2$ ~0.3 eV higher in energy than the experimental silicane bands $S_1$ and $S_2$. Another difference is that $\Sigma_3$ is not observed in the ARPES data. $\Sigma_3$ is a half-filled band formed by $p_z$ orbitals originating mainly from Si atoms without H. These atoms belong to the silicane sub-lattice that is closest to the Ag(111) surface. In the experiment, the silicane is supported by the Ag(111) surface and a hybridization with Ag orbitals might be the reason why we do not observe the weakly dispersive $\Sigma_3$ band around the Fermi level.

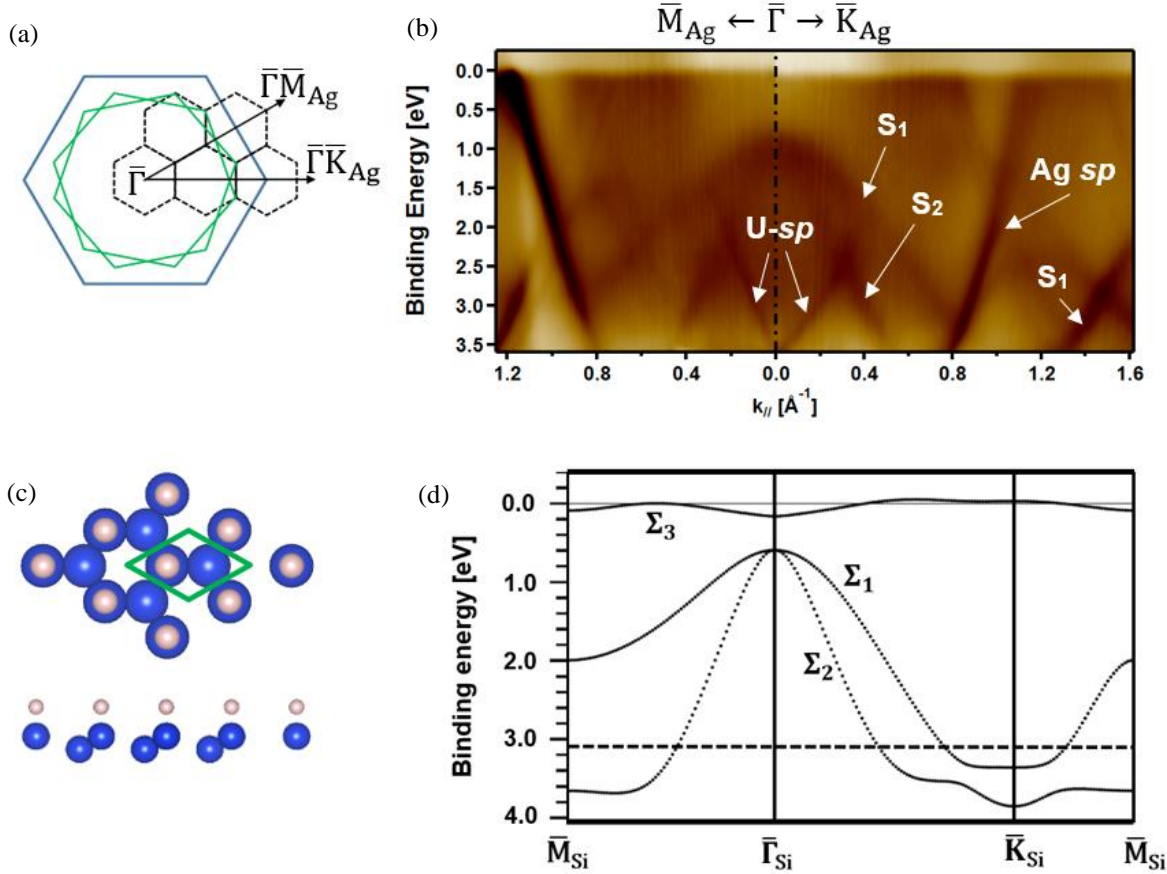

Fig. 3 (Color online) (a) SBZs of (1×1) Ag(111) (blue), (2√3×2√3) reconstruction (black) and two (1×1) silicane SBZs rotated by ± 10.9° (green). (b) ARPES data from silicane obtained at a photon energy of 26 eV along $\bar{\Gamma} \rightarrow \bar{M}_{Ag}$ and $\bar{\Gamma} \rightarrow \bar{K}_{Ag}$ of the Ag (1×1) SBZ. Note that these paths probe equivalent $k_{//}$ lines of the ±10.9° silicane SBZs. The energy range of the ARPES data extends down to a binding energy of 3.5 eV. Below that energy there is intense emission from the Ag 4d bands and no useful information of the silicene bands can be obtained. $S_1$ and $S_2$ are two dispersive bands which originate from silicane. The label Ag-*sp* indicates Ag bulk emission and U-*sp* points at dispersive features due to Umklapp scattering of the Ag-*sp* emission by (2√3×2√3) reciprocal lattice vectors. (c) Top and side view of a silicane model with hydrogen adsorbed on one of the two hexagonal sub-lattices of free-standing silicene. The green diamond represents the (1×1) silicane unit cell, blue and pink balls represent Si and H atoms, respectively. The bond length between Si and H atom is 1.52 Å, and the Δz of the two Si sub-lattices is 0.68 Å. (d) Calculated band structure of the silicane model in (c). The black dashed line indicates the binding energy limit of the ARPES measurement, relative to the top of the $\Sigma_1$ and $\Sigma_2$ bands at Γ.

In conclusion, our LEED and STM observations show clearly that the hydrogenation changes the structure of the ("2√3×2√3") silicene on Ag(111). The resulting structure corresponds to silicane which exhibits a hexagonal structure with a (1×1) periodicity with respect to the silicene lattice. In addition, the silicane layer shows a quasi-periodic (√7×√7)R19.1° vacancy pattern which we interpret as missing of Si atoms. The ARPES data from the hydrogenated silicene show two dispersive bands which we could unambiguously assign to silicane bands by the very good agreement with calculated band structure of the free standing silicane with H adsorbed on one side. This report of the electronic properties of silicane is important for future applications in Si based electronics.

*Acknowledgments*: Technical support from Dr. Johan Adell, Dr. Craig Polley and Dr. T. Balasubramanian at MAX-lab is gratefully acknowledged. The authors would like to thank Prof. Igor Abrikosov for discussion. Financial support was provided by the Swedish Research Council (Contracts No. 621-2010-3746, 621-2014-4764 and 621-2011-4426) and by the Linköping Linnaeus Initiative for Novel Functional Materials supported by the Swedish Research Council (Contract No. 2008-6582). The calculations were carried out at the National Supercomputer Centre (NSC), supported by the Swedish National Infrastructure for Computing (SNIC).

* Corresponding author: W. Wang, weiwa49@ifm.liu.se

Supplemental Material:

# Band structure of hydrogenated silicene on Ag(111): Evidence for silicane

W. Wang[*], W. Olovsson and R. I. G. Uhrberg

*Department of Physics, Chemistry, and Biology, Linköping University, S-581 83 Linköping, Sweden*

* Corresponding author: W. Wang, weiwa49@ifm.liu.se

In this supplementary section, we present results and analyses of silicane prepared by different exposures to atomic hydrogen. As mentioned in the paper, the exposures are quantified in terms of the number of Langmuirs (L) of $H_2$ molecules (1 L = $1\times10^{-6}$ Torr s). Figures S1(a), S1(b), and S1(c) show STM images of silicane prepared by exposures of 150, 200, and 400 L, respectively. The 150 L exposure corresponds closely to saturation coverage. The sequence of STM images in Fig. S1 shows the progress of Si desorption leading to areas of bare Ag(111) (dark areas in the STM images).

We also show ARPES data obtained along directions that correspond to the $\overline{\Gamma} - \overline{K} - \overline{M}$ and $\overline{\Gamma} - \overline{M} - \overline{\Gamma}$ lines of one of the two silicane orientations. These data give further details of the experimental band structure of silicane.

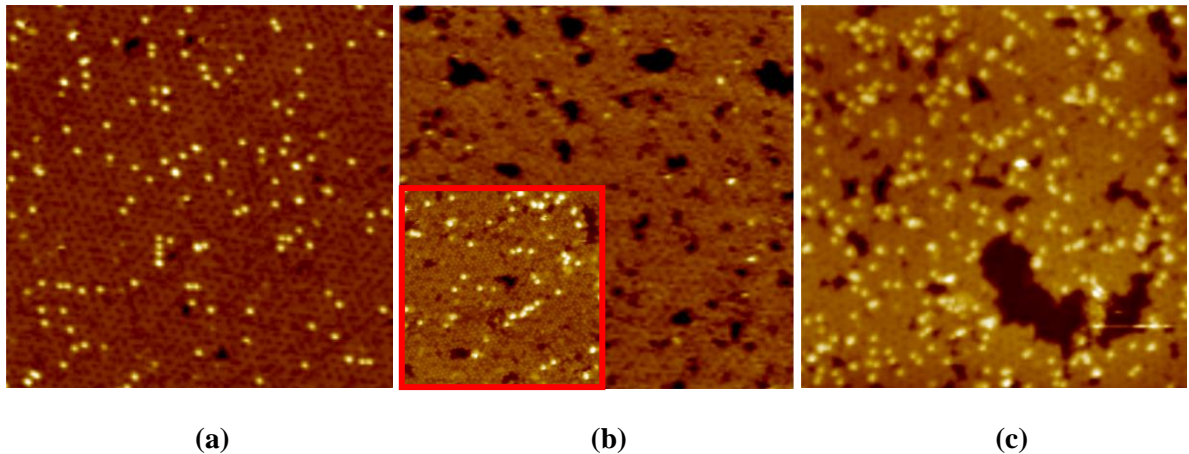

(a)　　　　　　　　　　(b)　　　　　　　　　　(c)

Figure S1. Filled state STM images of silicane (~40×40 nm$^2$) prepared by different exposures to atomic hydrogen (a) 150 L, (b) 200 L and (c) 400 L. The inset in (b) is a ~20×20 nm$^2$ STM image of higher quality to show the bright protrusions on the surface All images were recorded at room temperature in constant current mode with a tunneling current of 200 pA and a sample bias of -1.5 V.

By investigating the results of different exposures, we found that in our experimental setup, 150 L corresponds closely to a saturation exposure. We define this by the absence of the initial silicene structure, but with no significant areas of bare Ag (111), as in Fig. S1(a). The total area of the bright protrusions in Fig. S1(a) constitutes around 6-8 % of the imaged area which is close to what can estimated from Fig. 2(a) in Ref. [1]. Hence, we find it unlikely that they are due to contamination. Instead, we attribute them to unique features of hydrogenated ("2√3×2√3") silicene on Ag(111). A comparison of the STM images in Figs. S1(b) and S1(c) reveals the effects of extended exposures, 200 and 400 L, respectively. With increasing exposure, there is a larger number of holes down to the Ag(111) surface, some of which have developed into larger Si free areas as in Fig. S1(c). The bright protrusions and the Si free areas correspond to 18-20 % and 10-12 %, respectively in Fig. S1(c). We propose that the bright protrusions in the STM images are actually clusters formed by Si atoms detached from the silicane lattice. Since there are very few holes through the silicane layer in Fig. 1(a), it is reasonable to assume that the bright protrusions are made up of Si atoms which have been detached from the silicane layer resulting in the quasi-periodic vacancy structure. To summarize, Qiu *et al*. [1] interpreted the vacancies as missing

H atoms. We propose another possible explanation, i.e., the vacancies correspond to missing Si atoms. This would naturally explain the presence of the bright protrusions.

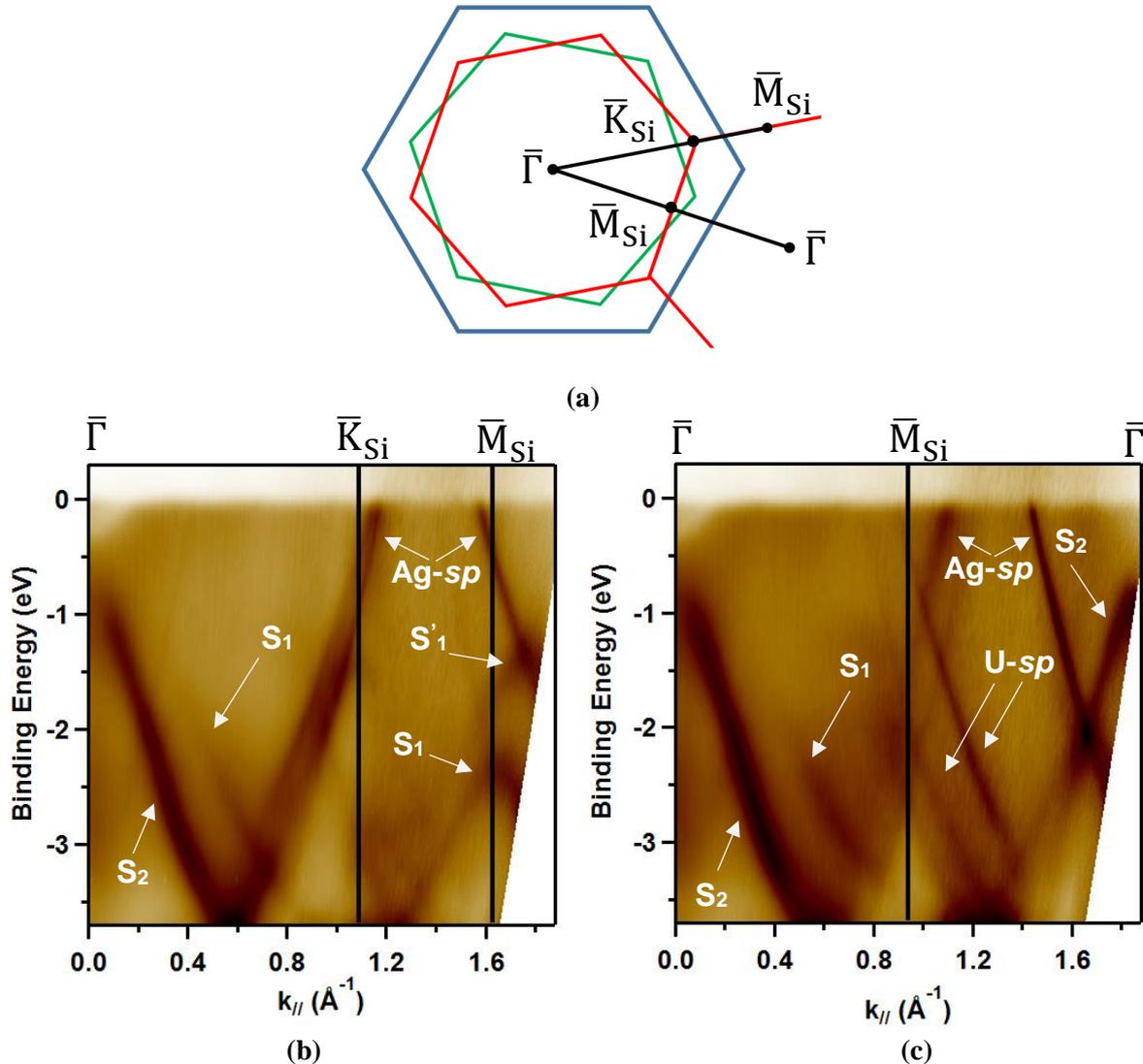

Fig. S2 (a) Schematic drawing of the 1×1 SBZs of Ag(111) (blue), and of the two ± 10.9° orientations of silicane (red and green). Black lines indicate the paths along which the ARPES data were recorded. For this choice, the $\bar{\Gamma} - \bar{K} - \bar{M}$ and $\bar{\Gamma} - \bar{M} - \bar{\Gamma}$ symmetry lines are probed for one of the silicane SBZs (red SBZ). (b) and (c) Dispersions of silicane bands, $S_1$ and $S_2$, obtained by ARPES at a photon energy of 19 eV. $S'_1$ corresponds to $S_1$ but comes from the non-symmetry line of the green SBZ. Ag-*sp* and U-*sp* correspond to Ag bulk emission and Umklapp scattered bulk emission, respectively.

In ARPES data measured using a photon energy of 19 eV the emission intensity of the $S_2$ band is significantly higher than for $S_1$ in the vicinity of normal emission. Near $\bar{M}_{Si}$ (at 1.63 Å$^{-1}$) in Fig. S2(b), there are two bands ($S_1$ and $S'_1$). We interpret them as $\Sigma_1$ bands from the two differently oriented silicane domains. When bands are mapped along the $\bar{\Gamma}\bar{M}_{Ag}$ direction [Fig. 3(b) in the paper], $S_1$ and $S'_1$ appear as one band since equivalent directions were probed for the two orientations. Furthermore, the turning point of the $S_1$ dispersion in Fig. S2(b) fits nicely with the $\bar{M}_{Si}$ value of 1.63 Å$^{-1}$. The intense band near the second $\bar{\Gamma}$ (1.88 Å$^{-1}$) in Fig. S2(c) corresponds to $S_2$ in the second SBZ. The binding energy difference between the top of the $S_1$ band at normal emission and at $\bar{M}_{Si}$ is ~1.6 eV which fits nicely to the calculated band structure of silicane in the manuscript.

Reference: [1] J. Qiu *et al.*, arXiv:1506.02134